\documentclass[prl,a4paper,superscriptaddress,twocolumn,amsmath,amssymb]{revtex4-1}

\usepackage{graphicx}% Include figure files
\usepackage{color}
\usepackage{bm}% bold math
\usepackage{dsfont}
\usepackage{epstopdf}
\usepackage{hyperref}
\usepackage{pdfpages}

\newcommand {\Fig}[1] {Fig.~\ref{#1}}
\begin{document}

%\title{Electrically-controlled quantum gates in donor nuclear spins in silicon}
\title{Conditional control of donor nuclear spins in silicon using Stark shifts}

\author{Gary Wolfowicz}
\email{gary.wolfowicz@materials.ox.ac.uk}
\affiliation{London Centre for Nanotechnology, University College London, London WC1H 0AH, UK}
\affiliation{Dept.\ of Materials, Oxford University, Oxford OX1 3PH, UK}  

\author{Matias Urdampilleta}
\affiliation{London Centre for Nanotechnology, University College London, London WC1H 0AH, UK}

\author{Mike L. W. Thewalt}
\affiliation{Dept.\ of Physics, Simon Fraser University, Burnaby, British Columbia V5A 1S6, Canada}

\author{Helge Riemann}
\author{Nikolai V. Abrosimov}
\affiliation{Institute for Crystal Growth, Max-Born Strasse 2, D-12489 Berlin, Germany}

\author{Peter Becker}
\affiliation{Physikalisch-Technische Bundesanstalt, D-38116 Braunschweig, Germany}

\author{Hans-Joachim Pohl}
\affiliation{Vitcon Projectconsult GmbH, 07745 Jena, Germany}

\author{John~J.~L.~Morton}
\email{jjl.morton@ucl.ac.uk}
\affiliation{London Centre for Nanotechnology, University College London, London WC1H 0AH, UK} 
\affiliation{Dept.\ of Electronic \& Electrical Engineering, University College London, London WC1E 7JE, UK} 

\date{\today}

%%%%%%%%%%%%%%%%%%%%%%%%%%%%%%%%%%%%%%%%%%%%%%%%%%%%%%%

\begin{abstract}
%One of the many challenges in designing quantum devices lies in balancing a degree of tunability against achieving long coherence times. For example, the small spin-orbit coupling of electron and nuclear spins of donors in silicon is one of the factors in giving them the longest measured coherence times of any solid state system, though it also limits the degree of electric field control achievable. 
%
Electric fields can be used to tune donor spins in silicon using the Stark shift, whereby the donor electron wave function is displaced by an electric field, modifying the hyperfine coupling between the electron spin and the donor nuclear spin. 
We present a technique based on dynamic decoupling of the electron spin to accurately determine the Stark shift, and illustrate this using antimony donors in isotopically purified silicon-28. We then demonstrate two different methods to use a DC electric field combined with an applied resonant radio-frequency (RF) field to conditionally control donor nuclear spins. The first method combines an electric-field induced conditional phase gate with standard RF pulses, and the second one simply detunes the spins off-resonance. Finally, we consider different strategies to reduce the effect of electric field inhomogeneities and obtain above 90\% process fidelities.
\end{abstract}

\maketitle

%%%%%%%%%%%%%%%%%%%%%%%%%%%%%%%%%%%%%%%%%%%%%%%%%%%%%%%
%\emph{Introduction- }
%One of the many challenges in designing quantum bits lies in balancing a degree of tunability against achieving long coherence times~\cite{XXX}. For example, the small spin-orbit coupling of electron and nuclear spins of donors in silicon is one of the factors in giving them the longest measured coherence times of any solid state system~\cite{XXX], though it also limits the degree of electric field control achievable.
 
The manipulation of donor spins in silicon is mainly realised by the application of resonant AC magnetic fields. These can be used to globally control a large ensemble of spins with high fidelity~\cite{Simmons2011}, and also to control a single spin~\cite{Pla2012}. However, applying local AC magnetic fields to donors in an interacting array is practically challenging, especially considering the number of high frequency, high power microwave lines in a mature device. For such reasons, the majority of scalable quantum computing architectures based on donors in silicon~\cite{Kane1998,Hill2005,Hollenberg2006}, combine globally applied AC magnetic fields with the ability to electrically tune the donor spin, via the Stark effect which primarily modulates the hyperfine interaction between the electron and nuclear spin. 

The Stark effect arises from a perturbation of the electron wave function as it is pulled away from the nucleus, mixing its ground state energy level with the higher orbital excited states \cite{Smit2008}. Following the original proposal by Kane~\cite{Kane1998}, several theoretical studies have examined Stark-tuning of both donor spins \cite{Friesen2005,Rahman2007b,Smit2008,Rahman2009}, while experiments have focused on measuring the Stark shift parameters using electron spin resonance (ESR) of the donor~\cite{Bradbury2006,Dreher2011a,Lo2014}. 
These measurements showed a typical Stark-shift induced change in the hyperfine coupling on the order of kHz (for electric fields around 0.1~V/$\mu$m). This falls well within the ESR linewidth (12~MHz in natural silicon, or typically 100~kHz in silicon-28 limited by the magnetic field homogeneity), making it impossible to Stark-shift the electron spin by more than a linewidth.
%as the Stark shift was always measured on the electron spin where it is challenging to resolve. This is partly due to the inherent unprotected nature of the electron spin: with it's large Zeeman interaction, it is quite sensitive to external magnetic field fluctuations that limit both the spin linewidth and coherence times. At reasonable electric fields (lower than the ionization energy), the Stark induced hyperfine shift is on the order of kHz, only 1-3 orders of magnitude faster than the decoherence rates of donor spins in isotopically enriched 28-silicon.

The nuclear spin, on the other hand, typically has much smaller spin resonance linewidth compared to the electron spin ($\gamma_{\rm n}/\gamma_{\rm e} = \sim4\times10^{-4}$, where $\gamma_{\rm n/e}$ is the gyromagnetic ratio of the nuclear/electron spin) and yet
the effect of the Stark-tuned hyperfine shift on the nuclear spin resonance frequency is nearly the same as for the electron spin.
As a consequence, using the Stark effect to tune spins in and out of resonance with globally applied fields becomes more readily achievable. Furthermore, the coherence times of the nuclear spin can reach minutes in ensembles \cite{Steger2012} and seconds even in single spin devices \cite{Muhonen2014}, allowing a much larger number of Stark-shift-controlled quantum operations to be applied. In this Letter, we examine the use of the Stark shift to conditionally manipulate the nuclear spin of antimony donors in silicon. We further demonstrate how the Stark shift can either be used as the basis for a simple controlled-phase gate or as a detuning gate to enable or disable the effect of a resonant magnetic radio frequency (RF) pulse. In the latter case, we are able to reach process fidelities above 90~\% fidelities.

%%%%%%%%%%%%%%%%%%%%%%%%%%%%%%%%%%%%%%%%%%%%%%%%%%%%%%%
%\emph{Materials and Methods- }
Measurements were conducted using an isotopically enriched silicon-28 float-zone crystal doped with antimony ($^{121}$Sb) at a concentration of $10^{14}$~cm$^{-3}$. Pulsed electron spin resonance (ESR) and electron-nuclear double resonance (ENDOR) experiments were realised in a Bruker X-band ($\approx$~0.3~T, 9.7~GHz) Elexsys system. The sample, 1.71~mm in thickness, was inserted between two metal plates connected to a semi-rigid copper nickel coaxial cable to apply the voltage pulses (see Supplementary Information), and sat in a continuous-flow helium cryostat at around 4.5~K. Voltage pulses up to 150~V, equivalent to 0.09~V/$\mu$m, were created with a low power high-voltage amplifier. The electric field was applied parallel to the static magnetic field.

\begin{figure}[t]%
\includegraphics[width=\columnwidth]{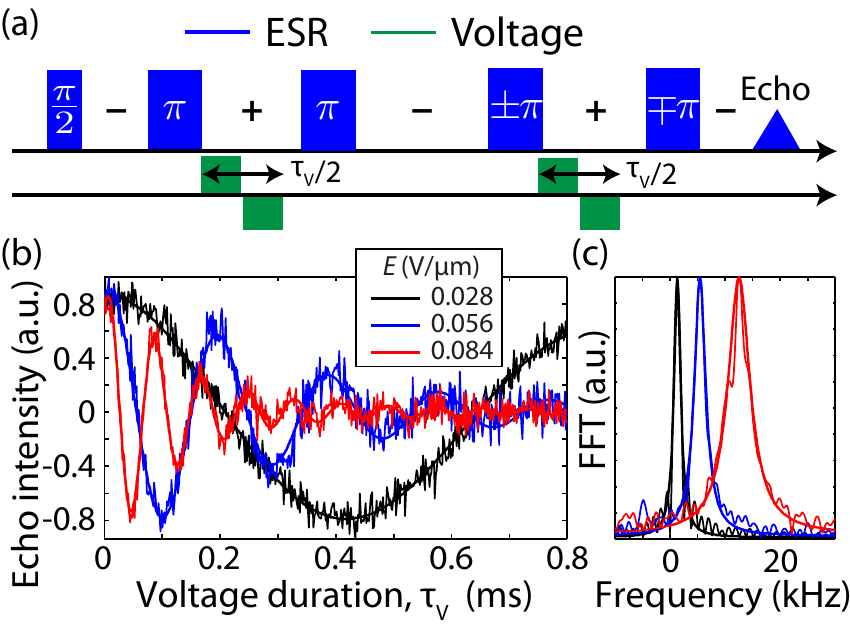}%
\caption{Measurement of the Stark shift in $^{28}$Si:$^{121}$Sb using dynamical decoupling.
(a) Uhrig Dynamical Decoupling (UDD) sequence with four refocusing pulses. As each $\pi$ pulse reverses the phase acquisition (see signs in sequence), the DC electric field is applied in between alternating pairs of $\pi$ pulses. Using bipolar (positive and negative) voltage pulses, the linear Stark shift contribution is eliminated and only the quadratic part remains. The last two ESR pulses were phase cycled to remove any stimulated echoes. 
(b) Electron spin phase evolution measuring using the $m_{\rm I} = -5/2$ ESR transition for different electric fields. (c) Fourier-transform showing the frequency shift distribution in the sample with a Lorentzian fit.
}
\label{fig:SampleCharact}%	
\end{figure}

%%%%%%%%%%%%%%%%%%%%%%%%%%%%%%%%%%%%%%%%%%%%%%%%%%%%%%%
%\emph{Sample characterization}
In order to measure the amplitude and distribution of the Stark shift in our system, we first confirm and then extend the measurements by Bradbury {\it et. al.} \cite{Bradbury2006} on the Sb donor electron spins. The Stark shift is measured as the electric sensitivity $\eta_{\rm A}$ and $\eta_{\rm g_e}$ of the hyperfine contact interaction $A$ and the electron g-factor $g_{\rm e}$, respectively ($|\gamma_{\rm e}| = g \mu_{\rm B}$, with $\mu_{\rm B}$ the Bohr magneton). Together, the application of an electric field $E$ results in a total frequency shift, for a particular electron or nuclear spin transition:
\begin{equation}
\Delta f(E) = \left[(\eta_{\rm A} A)\times\frac{df}{dA} + (\eta_{\rm g_{\rm e}} g_{\rm e})\times\frac{df}{dg_{\rm e}}\right]E^2
\label{eq:StarkShift}
\end{equation}
The derivatives $\frac{df}{dA}$ and $\frac{df}{dg_{\rm e}}$ are dependent on the static magnetic field, $B_0$, due to possible mixing between the electron and nuclear states, especially at low fields. In the high field limit, for an electron spin transition,  
$df/dA$ is equal to $m_{\rm I}$ (the nuclear spin projection) and $\frac{df}{dg_{\rm e}}$ is equal to $\mu_{\rm B} B_0$, while for a nuclear spin transition, $df/dA$ is equal to $1/2$ and $\frac{df}{dg_{\rm e}}$ is 0.

The frequency response is quadratic in the electric field, owing to the tetrahedral symmetry at each donor sites. However, any perturbation from this high-symmetry site (e.g. arising from local strains or internal electric fields from trapped charges in the crystal) gives an additional, linear Stark shift component. %Because this contribution originates strictly from inhomogeneities, it results in a signal decay in the experiments. 
The effect of this linear term can be suppressed by the application of alternating positive and negative voltage pulses as introduced by Bradbury et al.~\cite{Bradbury2006} and used here.

The Stark shift is measured in a Ramsey-type ESR experiment where the frequency shift $\Delta f(E)$ is acquired as a phase shift in the electron spin over time. In addition, refocusing pulses, such as used in a Hahn echo sequence, can significantly extend the acquisition time. The $T_2$ of the electron spin is 7~ms, limited by instantaneous diffusion \cite{Schweiger2001,Tyryshkin2003}, though additional magnetic field noise in our spectrometer reduces this time to about a millisecond. %This reduction appears as a global phase shift which can normally be corrected in ensemble experiments by looking at the magnitude of the signal \cite{Tyryshkin2003}. Here this phase error would prevent our Stark shift measurement, and as such is
To circumvent the effect of magnetic field noise, we use the UDD dynamical decoupling sequence \cite{Uhrig2007}, with voltage pulses applied between every other pair of decoupling microwave pulses (\Fig{fig:SampleCharact}(a)).

The phase evolution of the electron spin is shown in Fig. \ref{fig:SampleCharact}(b) as a function of the duration of the voltage pulse for various voltage amplitudes. For unipolar voltage pulses, the echo decays within one period of oscillation (see Supplementary Information) because of the inhomogenous nature of the linear Stark shift component. The application of bipolar voltage pulses can be used to select only the quadratic Stark shift, yielding many phase oscillations, limited only by the effective electric field distribution in the sample. Inhomogeneities in the electric field are expected due to surface roughness and imperfect alignment of the metal plate on the sample, however the Fourier transform of the signal (\Fig{fig:SampleCharact}(c)) shows a Lorentzian distribution which is suggestive of a different mechanism. This could arise from impact ionization events of donor from energetic free electrons under the electric field \cite{Kaiser1959}. Finally, the Stark shift parameters for $^{28}$Si:$^{121}$Sb where measured to be $\eta_{\rm A} = -3.54 \pm 0.11 \times 10^{-3} \mu$m$^2$/V$^2 $ and $\eta_{\rm g_e} = 5.3 \pm 1.9 \times 10^{-6}~\mu$m$^2$/V$^2$, in good agreement with published values~\cite{Bradbury2006} (see Supplementary Information).

\begin{figure}[t]%
\includegraphics[width=\columnwidth]{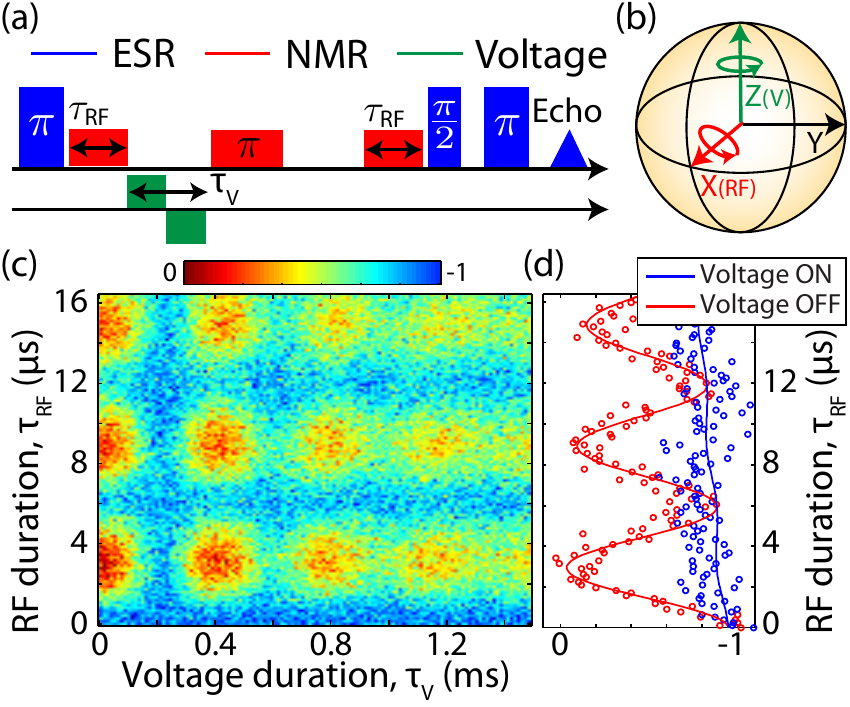}%
\caption{Electrically-controlled nuclear phase gate.
(a) Measurement pulse sequence where a Hahn echo is used to refocus the inhomogeneous broadening happening during the application of the voltage pulse.
(b) Bloch sphere representation of the nuclear spin rotations: in red, $X$-rotation from the RF pulse. In green, $Z$-rotation from the voltage pulse.
%The numbering gives the state position after each RF and voltage pulse. NOTE - I FIND THE NUMBERS A BIT CONFUSING, ARE THEY REALLY NECESSARY TO TELL THE STORY?
(c) The duration of the voltage and RF pulses are independently swept in a 2D map, resulting in spin rotations about the $X$- and $Z$-axes. The measured electron spin echo intensity is $-1$ for an unperturbed nuclear spin, and $0$ when the nuclear spin population is fully inverted, as is typical for Davies ENDOR measurements.
(d) Projection of the 2D map for $\tau_{\rm V} = 0$ ("OFF") and 0.2~ms  ("ON"), showing how the nuclear spin can made effectively insensitive to the applied RF field for an appropriate duration of the (150~V) voltage pulse.
}
\label{fig:NuclearPhase}%	
\end{figure}

%%%%%%%%%%%%%%%%%%%%%%%%%%%%%%%%%%%%%%%%%%%%%%%%%%%%%%%
%\emph{Nuclear Stark = phase gate}
We now move on to examine the effect of the electric field on the Sb nuclear spin, beginning with an analogous experiments to that above to create a nuclear phase gate. Measurements of the Sb nuclear spin are realised using a Davies ENDOR sequence (\Fig{fig:NuclearPhase}(a)) which projects the nuclear spin population onto the electron spin \cite{Davies1974,Tyryshkin2006}. 
Figure \ref{fig:NuclearPhase}(a-c) illustrates how combining the RF pulses with electric field control enables arbitrary X and Z rotations to be performed on the nuclear spin. This demonstrates that for specific values of the voltage pulse duration and amplitude, it is possible to enable or disable the effect of the applied RF pulses (\Fig{fig:NuclearPhase}(d)): when the voltage-induced nuclear spin phase shift is equal to $\pi$, the total sequence is always equivalent to a $\pi$ RF pulse, independent of the RF duration $\tau_{RF}$. The conditional operation on the nuclear spin had a total duration of about 0.5~ms which remains quite long even compared to the nuclear spin coherence times. However, the applied electric field of 0.09~V/$\mu$m is still far from the ionization energy of $\gtrsim1$V$/\rm \mu$m for donors in silicon \cite{Zurauskas1984,Friesen2005}, so in principle the voltage pulse durations could be significantly reduced.
%by up to two orders of magnitude to similar times as the RF pulses (7~$\mu s$ for a $\pi$ pulse here).
%WHERE DO YOU GET TWO ORDERS OF MAGNITUDE FROM AROUND 0.1V/UM TO 1V/UM?

%The duration of the voltage and RF pulses are independently swept in a 2D map. In blue when the ESR signal is -1, the nuclear spin population is inverted as is the case when $\tau_{RF}=0$. As the RF duration increases to $\pi/2$, the total operation becomes $2\pi$ corresponding to no ESR signal (red). When the voltage duration $\tau_V$ increases, the spins perform a Z rotation that will interfere with the X rotation, modifying the total trajectory.

%%%%%%%%%%%%%%%%%%%%%%%%%%%%%%%%%%%%%%%%%%%%%%%%%%%%%%%
%\emph{Nuclear Stark = off-res rabi}
We next use the electric field to tune the nuclear spin NMR frequency in or out of resonance with apply an RF field.
At our maximum electric field ($<$~0.1V/$\mu$m) and for the $m_{\rm I} = \pm5/2$ ESR transition, the frequency shift of 12~kHz cannot be resolved against the ESR linewidth of 50~kHz. For the nuclear spin (NMR transition) the frequency shift was measured to be 2.5~kHz, while the NMR linewidth is 500~Hz, two orders of magnitude smaller than for the electron (and partially limited by strain induced from the metal plates). 

\begin{figure}[t]%
\includegraphics[width=\columnwidth]{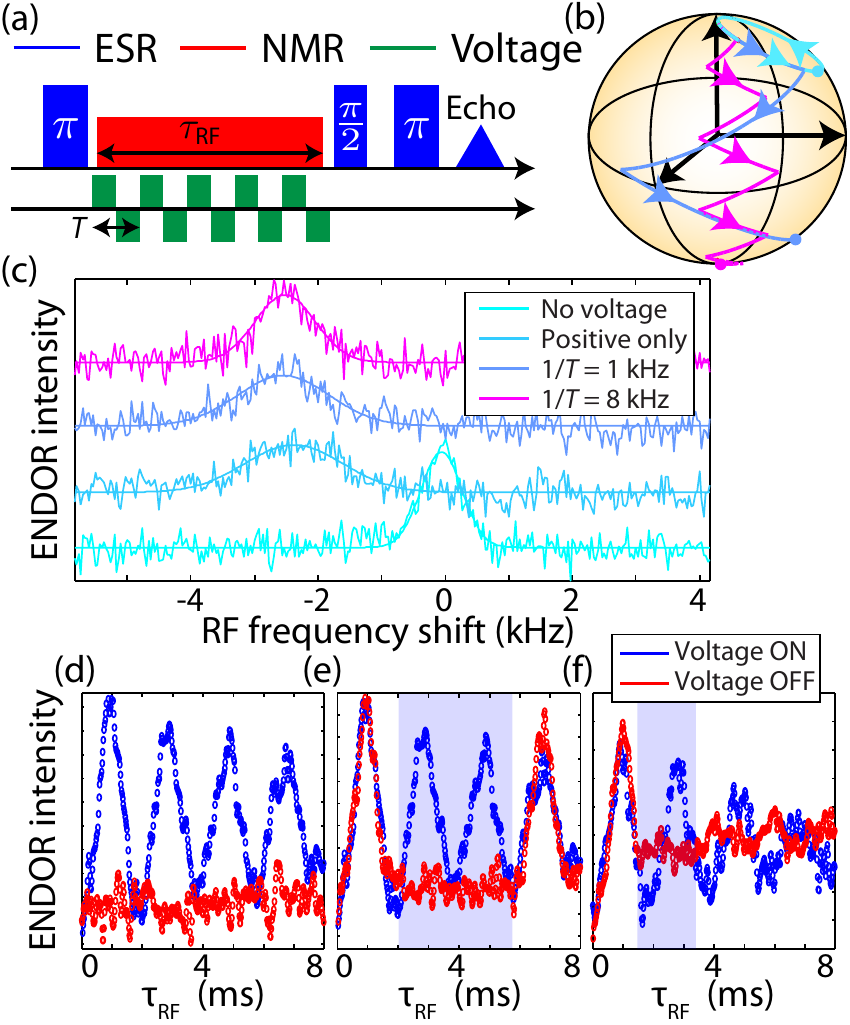}%
\caption{Electrically-tuning the nuclear spin transition frequency.
(a)~Davies ENDOR sequence with square wave voltage pulses.
(b)~Bloch sphere representation of the nuclear spin evolution in the rotating frame at the RF frequency which is resonant under a pure quadratic Stark shift. The linear Stark shift component adds additional off-resonant evolution, which can be compensated by applying a square-wave voltage pulse. 
(c)~ENDOR spectra measured with and without 150 V pulses, where $\tau_{\rm RF}$ = 1~ms corresponding to a $\pi$ pulse. 
For a single unipolar voltage pulse, the shifted line is broadened by the linear Stark shift. The shifted ENDOR line can be narrowed by increasing the frequency of a square wave bipolar voltage pulse, up to a point, when it becomes limited by electric field inhomogeneity.
(d-f)~Nuclear spin Rabi oscillations, where: (d) the voltage is constantly applied; (e) the voltage is applied only within the light blue window which starts when the nuclear spin is in an eigenstate; (f) the voltage is applied within the light blue window which starts when the nuclear spin is in a coherent superposition state.
}
\label{fig:NuclearRabi}%	
\end{figure}

Figure \ref{fig:NuclearRabi}(c) shows the ENDOR spectrum around the $m_{\rm S}=+1/2$ nuclear spin transition, recorded for different voltage pulses. 
Applying a unipolar voltage pulse for the full duration of the RF pulse is sufficient to fully shift the NMR line by more than a linewidth, however, the linewidth broadens due to electric field inhomogeneity and, in particular, the linear Stark effect. 
%The later is not a simple phase shift anymore, but results in a distribution of detuning and thus rotation axis $\sqrt{\Omega_R^2 + \delta\omega(E)^2}$, where $\Omega_R$ is the Rabi frequency of the nuclear spin and $\delta\omega$ the detuning from resonance. 
The latter can be suppressed using a voltage pulse of alternating polarity (i.e.\ a square wave with frequency of 8~kHz) as shown in figure \ref{fig:NuclearRabi}(a,b), allowing the Stark-shifted peak to narrow to a similar linewidth to the unshifted peak (see Supplementary Information).
%. This sequence prevent any unwanted evolution, similar to the Trotter approximation, due to the linear Stark effect: the peaks become narrower with higher frequencies until it stays constant . 
The electric field inhomogeneity is not corrected with this sequence and as a result the peak linewidth and intensity cannot be fully recovered.

The ability to tune Sb nuclear spins in and out of resonance with an applied RF field is shown in \Fig{fig:NuclearRabi}(d-f), where coherent rotation of the nuclear spins can be completely suppressed, or paused, for some given duration. This works effectively if the voltage pulse is applied when the nuclear spin is in an eigenstate, however, when the nuclear spin is in a coherent superposition, rapid dephasing caused by the applied electric field leads to incomplete recovery of the nuclear spin state after the end of the voltage pulse.
%
%This is due to the fact the nuclear spin superposition state acquires a phase of $2 \pi \Delta f(E) \tau_{RF}$ while the , and there distribution in $\Delta f(E)$as can be seen by the fact that the oscillation are off phase after the voltage pulse. This phase acquisition is a major issue as it must be tracked even for perfect pulses. 
%

\begin{figure}[t]%
\includegraphics[width=\columnwidth]{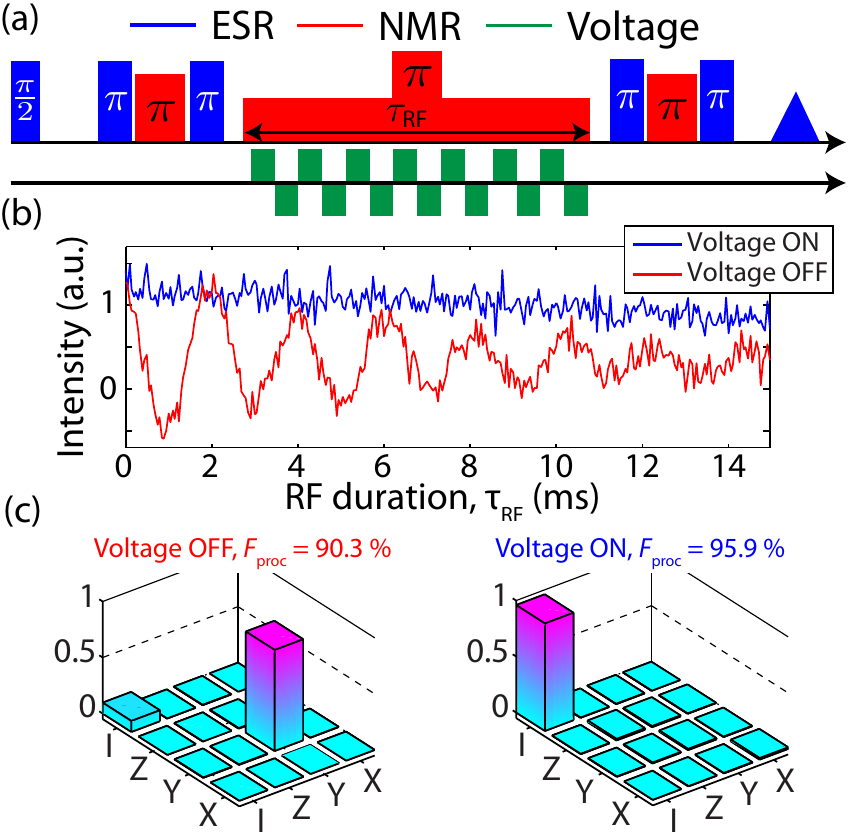}%
\caption{
(a) Sequence taken from \cite{Morton2008} used for transferring an electron state to a nuclear state. This is used to produce any input state ($\pm X$, $\pm Y$, $\pm Z$ and identity) due to limited RF phase control in our experiment. The $\pi$ RF gate in the middle is large bandwidth ($\approx$~30~kHz) and refocuses any dephasing when the voltage is on.
(b) Rabi oscillations with and without voltages for an input coherent superposition. The oscillations decay quite fast for negative amplitudes as they correspond to even number of $\pi$ pulses which prevent refocusing of the nuclear coherence.
(c) Quantum process tomography $\chi_{exp}$ matrices with voltage on (fidelity compared to a Y rotation) and voltage off (fidelity compared to the identity matrix). Only the real part is shown as the imaginary part is negligible.
}
\label{fig:Tomography}%	
\end{figure}

To solve this problem, a hard $\pi$ refocusing RF pulse can be applied half-way through the RF rotation (\Fig{fig:Tomography}(a)). It has the same frequency as the other RF pulses but with a much higher bandwidth (shorter, higher power pulse) that excites the nuclear spin regardless of any Stark-shift detuning. As shown in \Fig{fig:Tomography}(b), the recovered nuclear spin coherence now decays much more slowly.

With most of these source of errors suppressed, we finally evaluate the performance of the conditional nuclear spin gate using quantum process tomography~\cite{Childs2001,Jezek2003,O'Brien2004,Morton2008} (see Supplementary Information). In the absence of an applied voltage pulse, the gate indeed acts as a $\pi_Y$ rotation with a process fidelity $F_{\rm proc} = {\rm Tr}(\chi_{\rm exp}\chi_{\rm ideal}) = 90.3$~\%, where $\chi_{\rm exp}$ and $\chi_{\rm ideal}$ are the measured and expected process matrices, respectively (\Fig{fig:Tomography}(c)). The fidelity is limited here by the inhomogeneous broadening of the nuclear spin, $T_{\rm 2n}^*$. %Indeed, the $\pi$ pulse cancels out the broadband refocusing pulse and the total measurement is more akin to a nuclear Ramsey than a nuclear Hahn echo. 
Under an applied voltage pulse, the total gate resembles the identity operation with a fidelity $F_{\rm proc} = 95.9\%$.
% which is quite high for an ensemble measurement. 
%[LIMITED BY,ERROR BARS???]
%
%
%%%%%%%%%%%%%%%%%%%%%%%%%%%%%%%%%%%%%%%%%%%%%%%%%%%%%%%
%\emph{Outlook and Conclusions- }
Strategies for further increasing the gate fidelities and overcoming effects such as electric field inhomogeneities, including exploring composite pulses to address systematic errors~\cite{Wimperis1994}, or use of adiabatic sweeps to tune the spins through resonance during the microwave pulse~\cite{Tannus1997, Wu2013}. These techniques could be rather insensitive to variations in the electric field and microwave, though at the expense of longer gate durations. 

In conclusion, we have shown how the nuclear spin of a donor can be effectively controlled through a combination of RF excitation and an external electric field, either through electrically-controlled phase gates or by detuning spins off resonance. 
The techniques we discuss for dealing with electric field inhomogeneities could be particularly useful when using electric fields to address a subset of spins in a device, such as tuning out a row of spins in an array. 
By increasing the electric field closer to onset of donor ionization, these conditional nuclear spin gates could have timescales of order 10~$\mu$s, which compares favourable to the minutes-long nuclear coherence times. 
%%%%%%%%%%%%%%%%%%%%%%%%%%%%%%%%%%%%%%%%%%%%%%%%%%%%%%%

%\emph{Acknowledgements- }
We thank C.C.~Lo, A.M.~Tyryshkin and S.A.~Lyon for valuable discussions. This research is supported by the EPSRC through the Materials World Network (EP/I035536/1) and a DTA, as well as by the European Research Council under the European Community's Seventh Framework Programme (FP7/2007-2013) / ERC grant agreement no.\ 279781. J.J.L.M. is supported by the Royal Society. The $^{28}$Si-enriched samples used in this study were prepared from Avo28 material produced by the International Avogadro Coordination (IAC) Project (2004-2011) in cooperation among the BIPM, the INRIM (Italy), the IRMM (EU), the NMIA (Australia), the NMIJ (Japan), the NPL (UK), and the PTB (Germany). 

%%%%%%%%%%%%%%%%%%%%%%%%%%%%%%%%%%%%%%%%%%%%%%%%%%%%%%%
\bibliography{library}

%%%%%%%%%%%%%%%%%%%%%%%%%%%%%%%%%%%%%%%%%%%%%%%%%%%%%%%
%%%%%%%%%%%%%%%%%%%%%%%%%%%%%%%%%%%%%%%%%%%%%%%%%%%%%%%
%SUPPLEMENTARY INFORMATION
\pagebreak~\pagebreak
\includepdf[pages={1}]{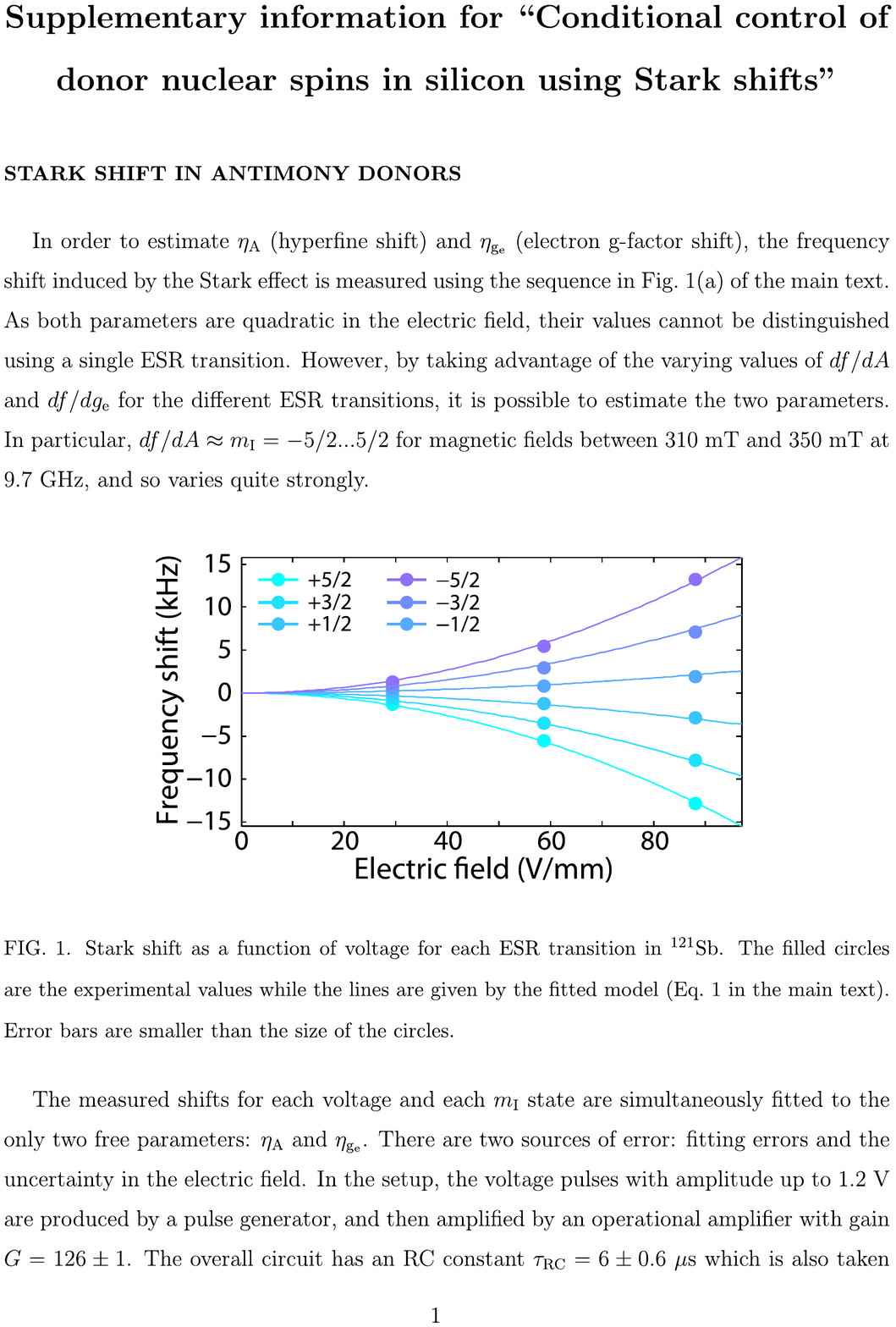}
~\pagebreak
\includepdf[pages={2}]{StarkENDORSOMArxiv.pdf}
~\pagebreak
\includepdf[pages={3}]{StarkENDORSOMArxiv.pdf}
~\pagebreak
\includepdf[pages={4}]{StarkENDORSOMArxiv.pdf}
~\pagebreak
\includepdf[pages={5}]{StarkENDORSOMArxiv.pdf}
~\pagebreak
\includepdf[pages={6}]{StarkENDORSOMArxiv.pdf}
~\pagebreak
\includepdf[pages={7}]{StarkENDORSOMArxiv.pdf}

%%%%%%%%%%%%%%%%%%%%%%%%%%%%%%%%%%%%%%%%%%%%%%%%%%%%%%%
%\emph{Author Contributions- }
%G.W., A.M.T., R.E.G, S.A.L., M.L.W.T and J.J.L.M. conceived and designed the experiments, G.W. and A.M.T performed the experiments, G.W., A.M.T., S.A.L. and J.J.L.M analysed the data, H.R., N.V.A, P.B, H.-J. P. and M.L.W.T provided materials, G.W. and J.J.L.M wrote the paper with input from all authors. 

\end{document}